\documentclass[manuscript]{acmart}

\usepackage{multirow}
\usepackage{array}
\newcolumntype{P}[1]{>{\arraybackslash}p{#1}}
\newcolumntype{M}[1]{>{\centering\arraybackslash}m{#1}}
\usepackage{subfigure}

\AtBeginDocument{%
  \providecommand\BibTeX{{%
    \normalfont B\kern-0.5em{\scshape i\kern-0.25em b}\kern-0.8em\TeX}}}

\copyrightyear{2021}
\acmYear{2021}
\setcopyright{acmcopyright}\acmConference[ARES 2021]{The 16th International Conference on Availability, Reliability and Security}{August 17--20, 2021}{Vienna, Austria}
\acmBooktitle{The 16th International Conference on Availability, Reliability and Security (ARES 2021), August 17--20, 2021, Vienna, Austria}
\acmPrice{15.00}
\acmDOI{10.1145/3465481.3470045}
\acmISBN{978-1-4503-9051-4/21/08}

\begin{document}

\title[Information Security Analysis in the Passenger-AV Interaction]{Information Security Analysis in the Passenger-Autonomous Vehicle Interaction}

\author{Mariia Bakhtina}
\email{mariia.bakhtina@ut.ee}
\author{Raimundas Matulevi{\v{c}}ius}
\email{raimundas.matulevicius@ut.ee}
\affiliation{%
  \institution{Institute of Computer Science, University of Tartu}
  \streetaddress{Narva mnt. 18}
  \city{Tartu}
  \country{Estonia}
  \postcode{51009}}
  

\begin{abstract}
  Autonomous vehicles (AV) are becoming a part of humans' everyday life. There are numerous pilot projects of driverless public buses; some car manufacturers deliver their premium-level automobiles with advanced self-driving features. Thus, assuring the security of a Passenger--Autonomous Vehicle interaction arises as an important research topic, as along with opportunities, new cybersecurity risks and challenges occur that potentially may threaten Passenger's privacy and safety on the roads. This study proposes an approach of the security requirements elicitation based on the developed threat model. Thus, information security risk management helps to fulfil one of the principles needed to protect data privacy - information security. We demonstrate the process of security requirements elicitation to mitigate arising security risks. The findings of the paper are case-oriented and are based on the literature review. They are applicable for AV system implementation used by ride-hailing service providers that enable supervisory AV control.
\end{abstract}

\ccsdesc[500]{Security and privacy~Domain-specific security and privacy architectures}
\ccsdesc[500]{Information systems~Process control systems}
\ccsdesc[500]{Software and its engineering~Risk management}

\keywords{autonomous vehicles, information system security risk management (ISSRM), human-computer interaction, threat modelling}

\maketitle

\section{Introduction}
An autonomous vehicle (AV) is defined as a system that can conduct dynamic driving tasks with the limited human intervention~\cite{sae2016taxonomy}. While creating a fully autonomous vehicle for everyday usage is still a big challenge, personal cars have advanced self-driving features implemented~\cite{SMofAS_Survey},~\cite{hagenziekerautomated}. 
Despite the autonomous vehicles' intelligence, a human is still kept on-the-loop and conduct supervisory control over the system and monitor the autonomous system. Thus with the increase of vehicle system automation, there is a concept switch from Human-Computer Interaction (HCI) to Human-Robot Interaction (HRI). Such a shift takes place as the nature of interaction design changes from control-oriented to supervisory control~\cite{HMIvsHRI}. However, according to~\cite{HCI_grandChal}
, the increase of  systems interconnectivity on which HRI is based leads to ``scalability issues making traditional security countermeasures inapplicable.'' That is why it is essential to define whether there is a difference between information security risk management (ISRM) in HCI and HRI, and whether the methods to treat ISRM in HCI can be applied to Human-AV interaction.

Security failure may cause vehicle damage, financial losses, disclosure of sensitive personal data, and road accidents~\cite{ENISA_SmartCars}. While much attention is paid to making autonomous driving ready-to-use technology, the urgency of security risk management is omitted. 
Infotainment systems that give a passenger understanding of the vehicle's behaviour and surroundings is considered as the one that presents the biggest attack potential for vehicle networks~\cite{hodge2019vehicle}. Numerous successfully implemented proof-of-concept attacks~\cite{ENISA_SmartCars} proves the possibility of remotely taking control of a vehicle's infotainment system and manipulating driving function. They demonstrate a lack of proper addressing security risks during the development. For the successful launch of autonomous vehicle projects, implementation of security measures and conducting risk management are required phases of AV system and supporting services lifecycles~\cite{ENISA_CAM2021}. 

One way to address this task is to separate AV's functionality into use cases and conduct risk analysis for each of them separately. The developed AV subsystems and their use cases should be reviewed from different perspectives to conduct an in-depth security risks analysis. One use case to be researched is the interaction of AV with the end-user - Passenger. This use case enables Passenger to get information about the AV functioning and conduct supervisory control over the vehicle. The case includes sharing personal and critical for safe transporting data between intelligent transportation system (ITS) components, e.g., AV controllers, a ride-hailing service provider, and the other infrastructure components.
This study aims to investigate \textbf{\textit{how information security risks in Passenger-Autonomous Vehicle interaction can be managed}}.

\begin{figure}[!ht]
    \centering
    \includegraphics[width=0.45\textwidth]{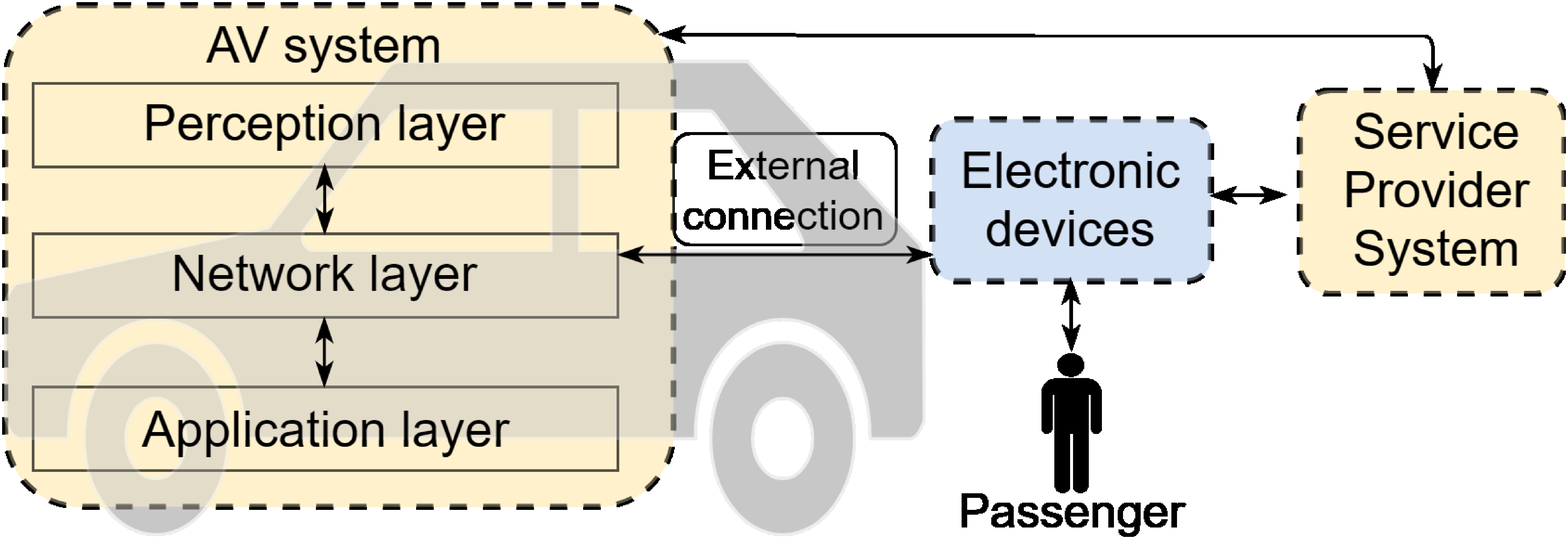}
    \caption{The researched autonomous vehicle ecosystem and its architecture~\cite{PassengerDataProtection}} \label{fig:AV_general_model}
    \Description{The researched ecosystem where Passenger-AV interaction takes place consists of the following elements: (i) AV system installed in the car - embraces components which compose perception, network and application layer; (ii) Service Provider System - information system (IS) used by ride-hailing service provider. Passenger uses electronic devices (e.g., personal on installed in the AV tablet or smartphone) for interaction with the systems. Electronic devices establish a connection with the AV system via an external connection.}
\end{figure}

In this paper, we propose an approach of security requirements elicitation based on the developed threat model, which was initially presented in~\cite{PassengerDataProtection} by one of the paper authors. The approach is created relying on the business process and the AV system architecture designed within the autonomous driving lab. The designed ecosystem is supposed to be used by a ride-hailing service provider to allow customers to use driverless ride-hailing services. As shown on Fig.~\ref{fig:AV_general_model}, the considered ITS incorporates (\textit{i}) AV system which is in charge of dynamic driving tasks of a single-vehicle, and (\textit{ii}) a service provider information system (IS) that offer infotainment service to passengers and help them to set interaction with a vehicle (later referred to as `Central System'). The systems can be either managed by the same or different service providers. It should be noted that a set of attacks on the in-vehicle network and central IS is still present but remains outside the scope of this study.

The paper is organised as follows: in Sect.~\ref{sec:CaseDescription}, we describe the case, followed by the background in Sect.~\ref{sec:Background} and the description of the research method in Sect.~\ref{sec:ResearchMethod}. Sect.~\ref{sec:CaseAnalysis} contains security analysis of the researched AV ecosystem.
Sec.~\ref{sec:Validation} introduces preliminary validation results of the elicited requirements. Finally, Sect.~\ref{sec:Discussion} concludes the paper by the discussion of the results, the lessons learned, and provides directions for the future research.

\section{Case Description} \label{sec:CaseDescription}

\begin{figure*}[!ht]
    \centering
    \includegraphics[width=1\textwidth]{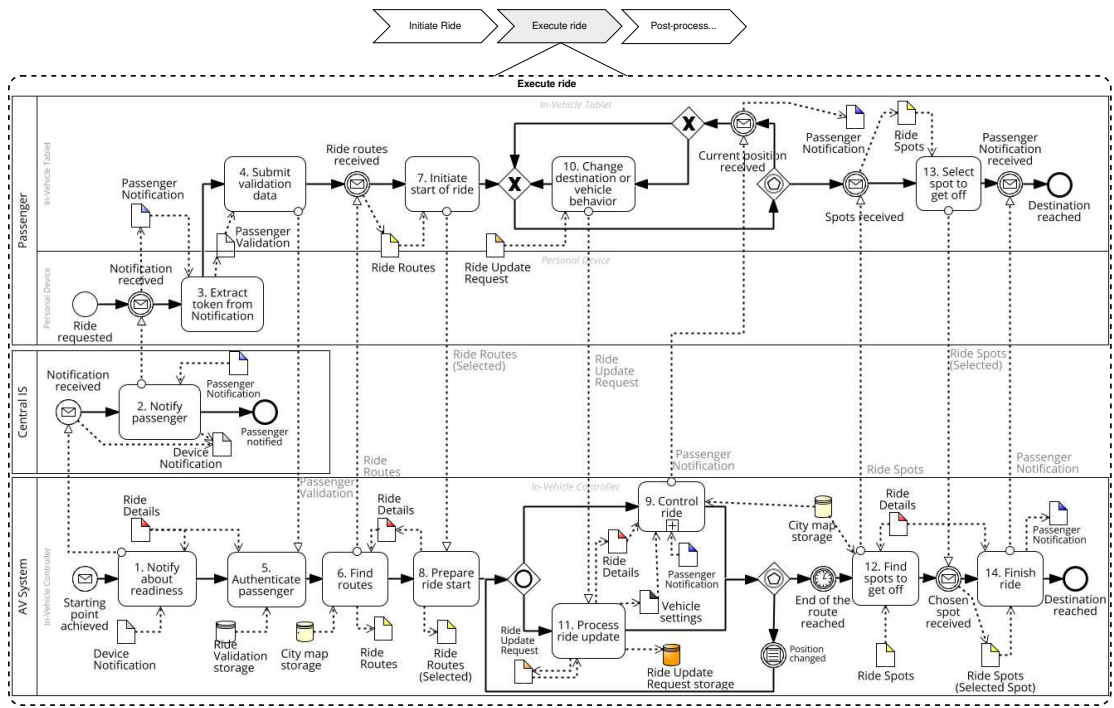}
    \caption{Ride Fulfilment business process~\cite{PassengerDataProtection}} \label{fig:RideFulfilmentsProcess}
    \Description{The Passenger-AV interaction is a part of the Ride Fulfilment process, consisting of three parts that deliver value to a Passenger, namely Ride Initiation, Ride Execution and Ride Post-Processing. Ride Initiation sub-process covers the handling of Passenger's request by a Central IS. It also covers the preparation of the AV system for the ride by conducting preliminary Ride Details processing and getting to the ride's starting point. The Ride Execution sub-process covers the phase of active Passenger-AV interaction where Central IS initially helps to set the interaction. Finally, Ride Post-processing takes places after Passenger left the vehicle. AV System should transfer Ride Update Request Storage copy to Central IS. Central IS finalises the ride by analysing the captured during the ride data to improve future services and archive Ride Detail to the storage to access them on demand.}
\end{figure*}

The Passenger-AV interaction occurs during the \textit{Ride Fulfilment} process, which consists of three parts that deliver value to a Passenger -- \textit{Ride Initiation}, \textit{Ride Execution} and \textit{Ride Post-Processing}. As the baseline of the surveyed processes, we used a user interface prototype 
designed in the autonomous driving lab\footnote{More information about the autonomous driving lab can be found at \textcolor{blue}{\url{https://www.cs.ut.ee/en/autonomous-driving-lab}}}. The prototype aims to increase trust in autonomous vehicles. We depict the Ride Fulfilment business process using Business Process Modeling Notation (BPMN) language to capture the process from the business perspective and show the data flows within it. The process model is presented in Fig.~\ref{fig:RideFulfilmentsProcess}. The analysis presented in this paper covers the second part of the described process – Ride Execution sub-process as only in this phase of \textit{Ride Fulfilment} a passenger actively interacts with the AV.
 
The Ride Fulfilment starts when a \textit{Passenger} initiates a ride by submitting a Ride Request in the Service Provider's App using a personal device. \textit{Central IS} processes the request and sends it to the \textit{AV System} of the assigned vehicle to execute the ride. When the AV achieved the starting point of the ride, \textit{Passenger} authenticates themselves and initiates the ride start. Once the ride started, \textit{AV system} controls the ride by executing dynamic driving tasks. Meanwhile, \textit{AV system}  informs \textit{Passenger} about the current location, and \textit{Passenger} can change the destination point or the vehicle behavior (e.g., speed) by making a request on the In-Vehicle Tablet on which the web client of the system is opened. When approaching the destination point, the vehicle asks \textit{Passenger} to select a spot to get off among the available spots near the destination. As soon as the selected spot achieved, \textit{Passenger} is notified and asked to leave the vehicle. Finally, \textit{Central IS} finishes the ride by processing the captured during the ride data to improve future services.

\section{Background} \label{sec:Background}

\subsection{Security Risk Management}

While such standards as ISO/IEC 2700x series, National Institute of Standards and Technology (NIST) special publications generally guide security risk management, the introduced methods, perspectives, and terminologies of security risk management vary from one standard to another. Depending on the risk analysis approach (quantitative or qualitative), the nature of the problem, and the analyst preferences, organisations are employing different security risk management methods \cite{ISSRM_alignment}, \cite{FundSecSysMod_RaimMat} like OCTAVE
, the NIST Cybersecurity Framework
,  MEHARI
. The domain model for information systems security risk management (ISSRM) has been developed~\cite{Dubois2010} to avoid misunderstanding between security experts and orchestrate standards mentioned above. According to the survey results \cite{ISRM_ISO27_Survey}, ISSRM was assessed as one of the most proficient concepts that implement ISO/IEC 27001 standard requirements. Security modelling languages support the ISSRM domain model \cite{ISSRM_alignment}, which helps cover the model's concepts using the corresponding tools. To analyse the selected case scenario, we are using the ISSRM domain model as the baseline.

\begin{figure*} [!ht] 
    \centering
    \includegraphics[width=0.7\textwidth]{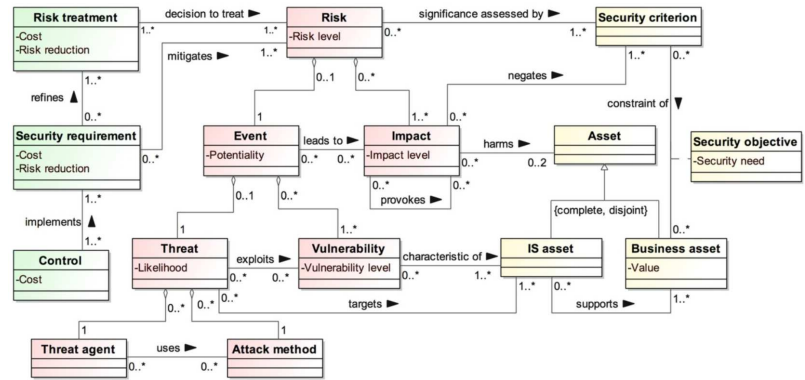}
    \caption{The ISSRM domain model~\cite{FundSecSysMod_RaimMat}: yellow entities represent asset-related concepts, red entities - risk-related concepts, and green entities - treatment-related concepts} \label{fig:fnDomainModel}
    \Description{The asset-related concepts describe the organisation's assets and their value to the organisation. The business asset represents any type of information vital for the proper processes flows and for achieving business needs. The (IS) assets support the defined business assets and are responsible for generating, manipulating, and storing new business assets. The security criteria of a business asset are defined by security objectives, which describe the security need of a system. The CIA (Confidentiality, Integrity, Availability) triad forms the main security criteria that can characterise business asset and should be delivered by the corresponding IS assets. The next group contains risk-related concepts, which correspond to risk itself and its constituent components. According to the model, risk is composed of a threat that exploits IS asset's one or few vulnerabilities, that leads to a harmful impact on the assets by negating the business asset's security criteria. In turn, a threat is an incident that is defined by a threat agent (also referred to as an adversary) who uses an attack method intending to target the selected IS asset. Finally, risk treatment-related concepts describe decision regarding risk treatment. Risk treatment is refined by the security requirements that mitigate one or more risks and which are implemented by means of controls.}
\end{figure*}

According to the ISSRM domain model (see Fig.~\ref{fig:fnDomainModel}), there are three key groups of concepts. The \textit{asset}-related concepts describe the organisation's assets, their value, and the reasoning why they should be protected. The \textit{risk}-related concepts correspond to risk itself and its components. The \textit{risk treatment}-related concepts describe how risks can be treated. 

\subsection{Related Work} \label{sec:RelatedWork}

Numerous studies have attempted to address information security and correspondent risk management in autonomous vehicles looking at the problem from different perspectives. Most of the researches, like~\cite{AVsecurity_AttacksDefTaxonomy_Thing_Wu}, \cite{CybSecChallengesInVehicularCom} and \cite{PotentialCyberAttacksOnAV}, are focusing on the security of in-vehicle components, vehicle-to-vehicle (V2V) and vehicle-to-infrastructure (V2I) communication. While AV-Human communication is mostly considered as interaction with pedestrians, the passenger's role is not widely discussed as well as the role of external service providers (e.g., ride-hailing service or ride-sharing). 

In ~\cite{PotentialCyberAttacksOnAV}, authors have discussed security and privacy threats in the case of autonomous and cooperative automated vehicles (CAVs) covering In-Vehicle, V2I, and V2V interaction. The study highlighted some attacks which can take place in the Passenger-AV interaction such as attacks targeting in-vehicle devices (e.g., hand-held devices connected to the infotainment system via USB, Wi-Fi or Bluetooth), other electronic devices and maps (used by a vehicle in case of non-real-time detection of the road). Mostly, authors reviewed the users' personal devices that help viruses and malware invade into the vehicle's electronics through the infotainment system and harm the in-vehicle network and its components functionality. Thus, considering autonomous vehicle security there are various attack vectors which researchers are addressing. With respect to it and the increase of AV technologies development, Thing and Wu in~\cite{AVsecurity_AttacksDefTaxonomy_Thing_Wu} proposed a comprehensive taxonomy of AV attacks and defenses that assist AV system architectures development. In~\cite{Parkinson_cyberThreats_CAVs} authors emphasize the increase of potential risks that affect or are conducted by the vehicle passengers as they have direct physical access to the system. Moreover, the research identified the following knowledge gap: ``it is unclear what personal data will be generated and stored, ... and what potential risks there are.''
In~\cite{KHAN2020} authors presented a taxonomy of threats and generalized attack surfaces for CAV applications. In the same paper, the role of the vehicle's components security within the CAVs supply chain was highlighted. For example, testing, anti-malware updates, and physical access to the AV components by original equipment manufacturers and vendors are recommended to be logged to preserve the security of the whole vehicle system and users' privacy. Additionally, human factor in the safety and security of CAVs was discussed. Our work is complimentary to the studies in~\cite{Parkinson_cyberThreats_CAVs, AVsecurity_AttacksDefTaxonomy_Thing_Wu, KHAN2020} as we study the scenario that covers security attacks which target vehicular network, vehicle system components and passenger's device, and propose measures to address corresponding security risks. The current work also embraces the scope of the mentioned works by illustrating their findings on a particular scenario.

Concerning the security risk management of AV systems, there exist several comprehensive guidelines that worth mentioning. The European Union Agency for Cybersecurity (ENISA) project~\cite{ENISA_SmartCars} considers the passengers of AV only in the context of how different attacks threaten the passenger's safety and pinpoints a need of raising awareness of passengers``with respect to security issues and how to prevent them, on a regular basis.'' In contrast, the guide~\cite{IPA_VehicleIS}, provided by Information-technology Promotion Agency (IPA), Japan, overviews the potential threats to autonomous vehicles on the high level of abstraction. It gives the general recommendations regarding security efforts in phases of automotive systems' lifecycle, which are not scenario oriented, but rather system functionality focused. However, they consider a passenger as a passive system user, which only obtains information from the infotainment system. In~\cite{hodge2019vehicle} authors defined in-vehicle infotainment systems as the one that presents the biggest attack potential for vehicle networks. Additionally, the mitigation techniques and procurement recommendations for infotainment systems which enables passengers' interaction with a vehicle was presented. Therefore, this work aims to build an analogue guideline with less focus on system functionality for the AV system developers and service-providers which use enable active Passenger-AV interaction.

 In the latest report~\cite{ENISA_CAM2021} ENISA highlighted the need for security risk management over the products and services lifecycles in the sector of connected and automated mobility, which includes an ecosystem of services, operations and infrastructure autonomous and connected vehicles. They also recommend conducting threat modelling for revealing relevant threat scenarios and address security issues in the early stages of system development. Additionally, the high level of the interconnectivity of the ecosystem components means that lack of security protection may lead to compromising the system at the scale of a fleet of vehicles. Thus, this paper complements the high-level recommendations in~\cite{ENISA_CAM2021} as we propose the security risks management approach to the scenario where the perimeters of ecosystem components intersect so that security measures and risk management should be considered holistically for preserving AV ecosystem protection.

To sum up, previous works discussed either general attack vectors on autonomous vehicle systems and defenses rather than examples of their applicability for the systems, or the general guides of security risk management of vehicular system, while none of the studies comprises a comprehensive overview of managing information security risks on a scenario level. Therefore, this paper aims to illustrate the the first stages of information risk management by incorporating a more technical and detailed attack methods discovery, and higher-level risk management approaches.

\section{Research Method} \label{sec:ResearchMethod}

The paper demonstrates the case analysis of the scenario, presented in Sect.~\ref{sec:CaseDescription}. This research aims to investigate \textit{how to manage information security risks in Passenger-Autonomous Vehicle interaction}. We are using the ISSRM domain model as a baseline for guiding the security risks and requirements definitions to address the main research question, therefore the sub-questions correspond to the domain model groups: 

\textbf{RQ1}: \textit{What assets should be protected in the Passenger--AV interaction?}

\textbf{RQ2}: \textit{What are the security threats in the Passenger--AV interaction?} 

\textbf{RQ3}: \textit{What are the security requirements to mitigate security threats in the Passenger--AV interaction?}

\begin{figure} [ht]
    \centering
    \includegraphics[width=0.45\textwidth]{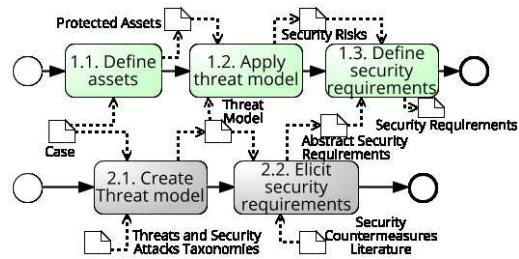}
    \caption{The research conduction map~\cite{PassengerDataProtection}} \label{fig:ResearchMethodProcess}
    \Description{The research of security risk management in Passenger-AV interaction consists of two parallel processes: theoretical artefacts development based on the literature review and the case analysis that illustrates the derived artefacts' application.}
\end{figure}

The research process is depicted in Fig.~\ref{fig:ResearchMethodProcess}. Two parallel processes are conducted: theoretical artefacts development based on the literature review and the case analysis that illustrates the derived artefacts' application.
Thereby, the paper demonstrates the results of the applied exploratory research of the under-researched case of Passenger--AV interaction.

The activity \textit{1.1.} answers the question \textit{RQ1} by defining the assets that should be protected based on the designed case. The business assets are extracted from the data structure of the researched system. For that, a class model is created using unified modeling language (UML) to define the key data entities and their dependencies. The combination of the flows depicted in the business process diagram and the data structure enables us to identify which assets should be protected, which security criteria assured and how assets are interconnected (see Sec.~\ref{sec:protectedAssets}). 

For answering the question \textit{RQ2}, the threat-driven approach is chosen for defining the risks in the research scenario. The attack libraries, taxonomies and the threat modeling framework are employed to determine a threat model on the step \textit{2.1} (see Sect.~\ref{sec:ThreatModel}). After, the threat model is applied to the defined protected assets (activity \textit{1.2}), resulting in the developed risks. 
Security countermeasures should be defined to mitigate the risks. Thereby, Sect.~\ref{sec:SecReqs} answers the question \textit{RQ3} demonstrating the security requirements elicitation (activity \textit{2.2}) based on the security countermeasures literature.

\section{Security Risk Management} \label{sec:CaseAnalysis}

This section aims to provide readers with the necessary background about threat modelling, including common threats taxonomies and libraries from which a threat model for Passenger-AV interaction is developed. Also the section demonstrates the results of security risks analysis for the presented in Section~\ref{sec:CaseDescription} scenario that results in the derived the threat model and security risks. Finally, the possible security measures to address risks are presented in the form of security requirements. 

\subsection{Protected Assets} \label{sec:protectedAssets}

This section answers the \textbf{RQ1} by defining the assets that should be protected. In the Passenger-AV interaction, \textit{business assets} (BA) are presented by transmitted data, vital for the proper processes flows. The \textit{system assets} support these business assets and are responsible for generating, manipulating, and storing new BAs. The \textit{security criteria} of a business asset are defined by security objectives, which describe the security need of a system. Confidentiality, integrity, and availability, also known as CIA triad, forms the main security criteria which can characterise business assets.

\begin{figure*} [ht]
    \centering
    \includegraphics[width=0.8\textwidth]{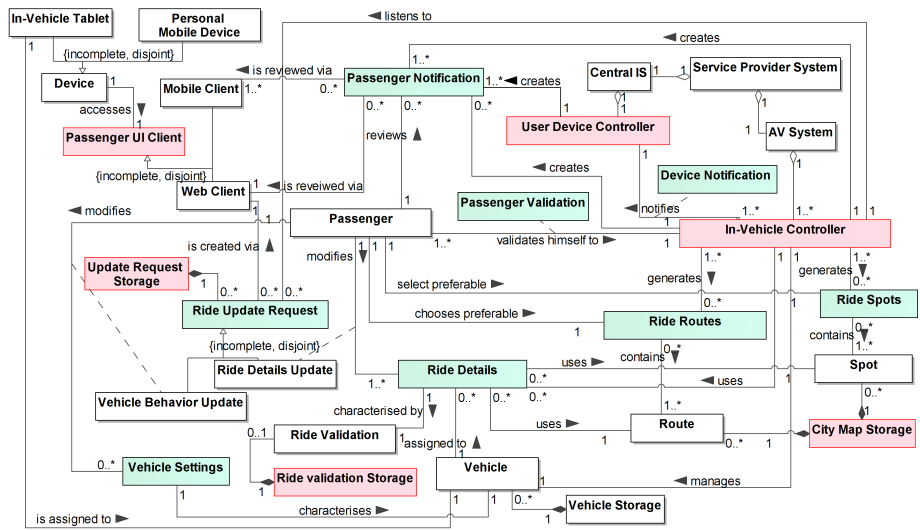}
    \caption[Data structure of the system]{Data structure of the system~\cite{PassengerDataProtection} (green entities - business assets; red entities - system assets)} \label{fig:fnDataStr}
    \Description{The data structure is depicted in the form of a UML class diagram. The identified business assets are illustrated on the diagram as green entities, and system assets are depicted as red entities.}
\end{figure*}

The general data structure of the system is depicted in Fig.~\ref{fig:fnDataStr} in the form of a UML class diagram. The identified business assets are illustrated on the diagram as green entities, and system assets are depicted as red entities.

As it case be seen from Fig.~\ref{fig:RideFulfilmentsProcess} and \ref{fig:fnDataStr}, \textit{Ride Details} is a central asset used by \textit{In-Vehicle Controller} for the ride execution. This entity contains such fields as a \textit{starting point}, and a \textit{destination} of the ride, collects information about \textit{selected routes}, \textit{ride spots} to get off, an involved in the ride \textit{vehicle}. Also, it contains the reference to the entity, which corresponds to \textit{Passenger} and stores her personal data, e.g. payment details. Meanwhile, \textit{Passenger Validation} asset contains credentials which a Passenger should use for starting a dialog with the system, and, consequently, start the ride. As a result, the \textit{Ride Details} asset aggregates all the other assets to enable Passenger conduct supervisory control over the AV during Ride Fulfilment.

Now let us consider how the system assets are organised. A service provider system consists of the two main components: Central IS and AV System.  \textit{In-Vehicle Controller} represents the back-end part of the AV system, and it is in charge of accessing data storage, conduction most of the calculation, and data manipulation functions. \textit{User-Device Controller} is a back-end part of Central IS. \textit{Passenger UI Client} in the observed system represents the front-end part and is separated into \textit{Web Client} and \textit{Mobile Client}, which corresponds to the front-end part of AV System and Central IS, respectively. Thus AV system communicates with a Passenger via Web Client opened on the In-Vehicle Tablet, while Central IS interacts with a Passenger via a mobile app installed on the Personal Mobile Device. Other components of the architecture is application programming interfaces (APIs), which facilitate communication between the system components - \textit{Central IS API} and \textit{AV System API}. 

\subsection{Threat Model} \label{sec:ThreatModel}

According to ~\cite{ThreatModeling_Shostack} and~\cite{FundSecSysMod_RaimMat}, information security risks are mostly defined by the attacks that an adversary employs to target a system assets. Thus, threat-driven approach for risks identification is commonly used~\cite{InfoSec_4thEdition, ThreatModeling_Shostack} for guiding the ISRM. The threat model should be defined as a primary step for risks identification. This subsection reviews the common threat taxonomies, and attack libraries focused on the attacker's tactics, techniques, and procedures (TTP)~\cite{CyberThreatModSurvey}. The primary deliverable of this subsection is a threat model for the Passenger-AV interaction scenario defined in Section~\ref{sec:CaseDescription}. 

\subsubsection{Threat taxonomies} 

The selected four resources of the threats and attacks were chosen due to the following reasons: (\textit{i}) 
the repositories are enterprise-neutral and technically focused as they do not put any limitations on a specific enterprise, its architecture, or assets but instead concerned with the overall technological environment; (\textit{ii}) the threats and attacks within the repositories are described in details, illustrated by real case implementations and the attacks supported by high-level mitigations; (\textit{iii}) the classified attacks and threats are relevant for the researched system architecture and process; and, finally (\textit{iv}) the combination of the repositories, taxonomies enable to derive threats for the system starting from the general attack vectors until the concrete threats covering the full scenario.
The \textit{STRIDE} approach~\cite{Stride} is designed for eliciting system security threats. STRIDE is supposed to be used at the beginning of ISRM during the defining potential risks and attack vectors. 
The Common Attack Pattern Enumeration and Classification~\cite{Capec} (\textit{CAPEC}) is a comprehensive, community-created catalog of attack patterns. It defines the informal taxonomy of attack-pattern classes and provides the formal description of each attack class. The taxonomy is organised hierarchically based on its domain and mechanisms of attack specifying the vulnerabilities it addresses. CAPEC is supported by references to the targeted vulnerabilities and possible mitigations.
Adversarial Tactics, Techniques, and Common Knowledge (\textit{ATT\&CK}) framework~\cite{Mitre_ATT_CK} is a knowledge base of adversarial techniques which helps to classify attacker's actions for different platforms (e.g., Windows, Android). 
It is focused on techniques in the context of tactics an adversary wants to apply to attack a specific component or endpoint. Concrete procedure examples support each technique an adversary may use, system requirements for implementing the tactics, possible detection methods, and mitigations. The techniques are mapped to the corresponding attack patterns. Another resource for the threat model creation is the list of vulnerabilities provided by the Open Web Application Security Project (\textit{OWASP})~\cite{OWASP_webPage} is considered a starting point for developing secure software focused on defensive mechanisms and controls.
The approach does not consider the prospect of a threat agent or any application implementation details.  
Thus, for each specific case, a threat agent, assets, and corresponding impact should be considered besides, respectively.

\subsubsection{The threat model for Passenger-AV interaction} 

\begin{figure*} [h]
    \centering
    \includegraphics[width=0.77\textwidth]{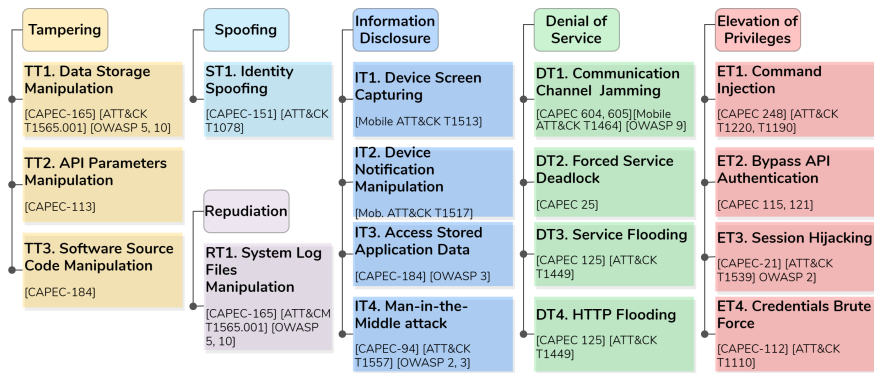}
    \caption{The threat model for the Passenger--AV interaction~\cite{PassengerDataProtection}} \label{fig:fnThreatMap}
    \Description{The proposed threat model is based on the attacks and system vulnerabilities from CAPEC, ATT\%CK, OWASP Top Ten taxonomies and libraries. The categories from the STRIDE approach group the threats. The proposed threat model is supposed to be used as a baseline for security risks identification and further security requirements elicitation.}
\end{figure*}

Fig.~\ref{fig:fnThreatMap} illustrates the derived threat model for the Passenger--AV interaction. The model contains 17 threats that an adversary can exploit during the Ride Execution process. The threats are organised into six groups. Each threat is supported by the reference to the source it was elaborated from. The detailed description of the threats (targeted vulnerability, threats agent, attack method, and potential impact) can be found in~\cite{PassengerDataProtection}.

\textit{\textbf{Spoofing}} refers to identity spoofing attacks where an attacker pretends to be a legitimate passenger. To violate the system's \textit{authentication} mechanism, an attacker uses the obtained credentials (ST1). In the case of \textit{\textbf{tampering}}, an attacker intentionally modifies a system, network, its behavior, or the data to violate their \textit{integrity}. These threats target data storage (TT1) and software source code files (TT3), which are used during the ride execution and are critical to the general trip safety and data reliability. As the Passenger-AV interaction includes communication between few separate entities (AV system and Central IS), API parameters can be manipulated for changing the normal entities communication (TT2). \textit{\textbf{Repudiation}} attacks are targeting the business layer during which the system cannot track and log actions accurately. As a result, the system claims that the activities were not done even if they were, or vice versa. By manipulating the system log files that keep track of both passenger's activities during the ride and the data about the driving task execution, an attacker can influence the current and the future ride that uses the historical data (RT1). \textit{\textbf{Information Disclosure}} groups the threats in which the confidentiality of the data is violated by providing access to it to someone who is not supposed to have access. It refers to accessing data while it is stored locally (IT3), displayed on the mobile device to a passenger (IT1, IT2), or in the transmission between systems or their components (IT4). Such attacks intend to gather information required for further attacks. \textit{\textbf{Denial of Service}} attacks are focused on consuming resources needed to provide service to a Passenger, and as a consequence, the \textit{availability} of the information is violated. The threats target either the communication channels' resources (DT1 and DT4) or computational resources (DT2 and DT3). \textit{\textbf{Elevation of Privileges}} threats refer to allowing an attacker to have authorisation permissions that he was not supposed to have, thereby violating the system's \textit{authorization}. It can be achieved either using the obtained legitimate credentials (ET3 and ET4) or by a more sophisticated manual bypassing the existing authentication mechanisms (ET1 and ET2). It should be noted that the identified threats are interconnected as implementation of one of them  enables execution of another. For example, successfully implemented\textit{ IT4. Man-in-the-Middle attack} enables execution of \textit{TT2. API Parameters Manipulation} which in turn may result in \textit{DT2. Forced Service Deadlock}.

\subsection{Security Risks} \label{sec:SecRisks}

\begin{figure*} [htbp]
\begin{tabular}{cc}
  \begin{minipage}{0.7\textwidth}
        \includegraphics[width=1\textwidth]{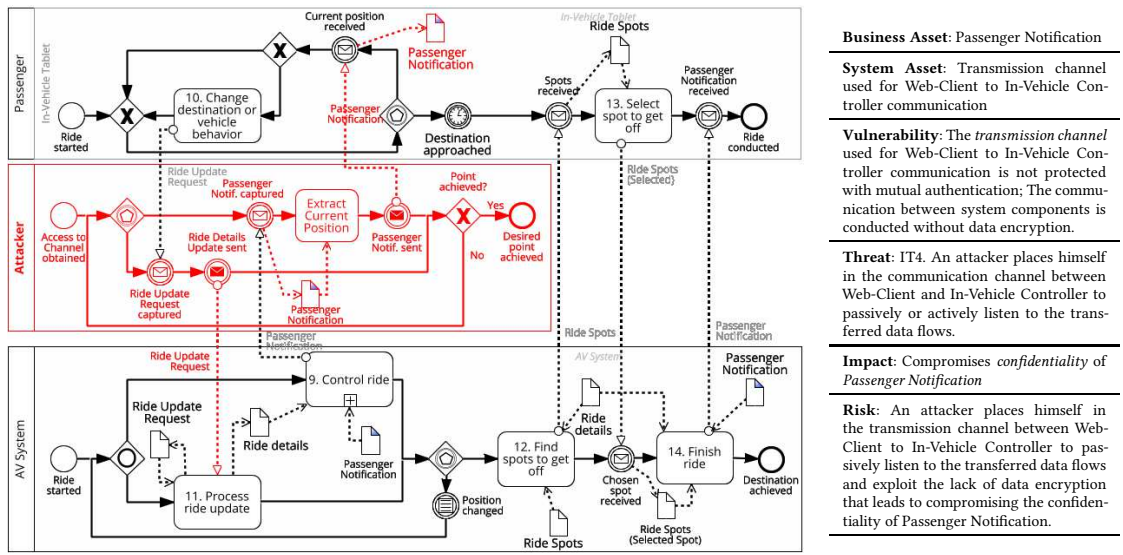}
    \end{minipage}
&
  \begin{minipage}{0.3\textwidth}
        \begin{tabular}{p{3.5cm}}
            \scriptsize{\textbf{Business Asset}: Passenger Notification}\\ \hline
 
            \scriptsize{\textbf{System Asset}: Transmission channel used for Web-Client to In-Vehicle Controller communication}\\ \hline
    
            \scriptsize{\textbf{Vulnerability}: The \textit{transmission channel} used for Web-Client to In-Vehicle Controller communication is not protected with mutual authentication; The communication between system components is conducted without data encryption.}\\ \hline
 
            \scriptsize{\textbf{Threat}: IT4. An attacker places himself in the communication channel between Web-Client and In-Vehicle Controller to passively or actively listen to the transferred data flows.} \\ \hline
 
            \scriptsize{\textbf{Impact}: Compromises \textit{confidentiality} of \textit{Passenger Notification}}\\ \hline
 
            \scriptsize{\textbf{Risk}: An attacker places himself in the transmission channel between Web-Client to In-Vehicle Controller to passively listen to the transferred data flows and exploit the lack of data encryption that leads to compromising the confidentiality of Passenger Notification.}\\ \hline
        \end{tabular}
    \end{minipage}
\end{tabular}
    \caption{Risk IR5: Man-in-the-Middle attack~\cite{PassengerDataProtection}} \label{fig:fnIR5MitM} 
    \Description{To illustrate the attack implementation, we are using the security extension to BPMN, which supports the ISSRM domain model. The diagram shows an attacker as an additional entity that intercepts in the transmission channel aiming to define when the AV with the Passenger reaches the desired place on their route. The figure also contains the components of the risk - business asset and supportive system asset, the targeted vulnerability,threat, its impact on the business asset.}
\end{figure*}

The current subsection answers \textbf{RQ2} by identifying security risks based on the developed threat model for Passenger-AV interaction.
The risk model for the observed scenario can be derived by instantiating attacks from the threat model to the business assets and its vulnerabilities. As a result, for the assets identified in Sec.~\ref{sec:protectedAssets}, the risk model includes 22 information security risks that can take place in the Service Provider System. The complete model can be found in~\cite{PassengerDataProtection}. Among the derived risks 13 risks are targeting Passenger Notification, and 8 out of 22 risks are targeting confidentiality of Passenger Notifications. Furthermore, some risks includes the harm to the system components, which as a result may result in getting access to any sensitive data which is visible to the system. To illustrate the attack implementation, we are using the security extension to BPMN~\cite{BPMNExt}, which supports the ISSRM domain model. Fig.~\ref{fig:fnIR5MitM} contains an example of the derived security risks -- namely, IR5 the Man-in-the-Middle (MitM) attack execution which targets \textit{Passenger Notification}. According to CAPEC, the MitM attack required medium skills level required, but has high impact, as it enables an adversary to conduct further attacks on the system. In Fig.~\ref{fig:fnIR5MitM}, we see an attacker as an additional entity that intercepts in the transmission channel aiming to define when the AV with the Passenger reaches the desired place on their route.

The implementation of Man-in-the-Middle attack (threat IT4) primarily aims to negate the confidentiality of Passenger Notification. However, the effective delivery enables an attacker to conduct a set of further attacks that already may target the vehicle's functions, which may provoke the loss of passenger's safety. Therefore, during the risk assessment and further decision about its treatment, it is vital to consider the direct impact and the impact of the risks that may be provoked by it — consequently, a threat-driven approach helps make requirements prioritisation. A higher priority should be given to the requirements that prevent risks with the highest impact and which stop an attacker from conducting further attacks.

\subsection{Security Requirements} \label{sec:SecReqs}

The current subsection contains a list of the derived security requirements that answers the last research question (\textbf{RQ3}).

\subsubsection{Security requirements elicitation}
According to~\cite{SecReqsFiresmith} and~\cite{FundSecSysMod_RaimMat}, a \textit{security requirement} is a condition of the domain environment that should be met in order to mitigate one or more security risks and utilising security controls implemented in the system.
In \cite{NHTSA}, National Highway Traffic Safety Administration (NHTSA) presents the recommended guideline to the automotive industry for the vehicle’s electronic architecture. It is intended to improve vehicle cybersecurity by implementing security controls. They also emphasise the necessity of using information technology security suite and standards (ISO 2700x series, CIS\cite{CISControls}). Similarly, the report by ENISA~\cite{ENISA_SmartCars} contains a set of good practices for smart cars. It stresses that for conducting information security management, it is important to use aforementioned standards along with SAE J3061~\cite{sae2016j3061} and NIST 800-53\cite{nist_sp800_63b}. IPA proposes the guide~\cite{IPA_VehicleIS} for achieving a security level in the automotive systems. They highlight the security management by implementing the security function design (in the sense of encryption, authentication, and access control), which should be enhanced with secure coding, security testing, and user training.

Meanwhile, threat-driven requirements elicitation approach supports security requirements categorisation: (i) \textit{preventive}; (ii) \textit{detective}; (iii) \textit{corrective}. The analogue taxonomy of security countermeasures for the AV defence is presented in  \cite{AVsecurity_AttacksDefTaxonomy_Thing_Wu}. Such requirements categorisation enables their prioritisation based on the impact of the risks. For example, a higher priority could be given to preventive requirements that preserve threats that enable further attacks execution. 

\subsubsection{Defined security requirements}
For mitigating the defined risks in the Passenger-AV scenario, we considered the security controls from the aforementioned standards and libraries. The requirements were later defined using the inductive approach from the found controls.
Using \cite{OWASP_webPage}, \cite{Capec}, \cite{CISControls}, \cite{nist_sp800_63b}, \cite{Mitre_ATT_CK} as the main primary sources, we have elicited 56 requirements which are supported with the possible implementations (i.e. security control components), organised in groups of the correspondent treated threats. The full list of elicited requirements can be found in~\cite{PassengerDataProtection}, while Table~\ref{tab:SecReqsIdentification} represents few examples.

\begin{table*}[htbp]
\caption{Security countermeasures identification (P - preventive, D - detective, C - corrective)}
\begin{center}
\begin{tabular}{|p{2.6cm}|p{0.4cm}|p{3.4cm}||p{2.4cm}|p{0.4cm}|p{3.4cm}|}
\hline
\scriptsize\textbf{Security Requirements}&\scriptsize\textbf{Class}&\bf\scriptsize{Security Control Components}&\scriptsize\textbf{Security Requirements}&\scriptsize\textbf{Class}&\bf\scriptsize{Security Control Components} \\
	\hline 
    \multicolumn{3}{|c||}{{\scriptsize{\textbf{IT2. Device Notification Manipulation.}}}}&  \multicolumn{3}{c|}{{\scriptsize{\textbf{DT3. Service Flooding.}}}}\\ \hline
	\scriptsize{\textbf{IT2.R1.} The system should protect sensitive data provided to the mobile app.}& \scriptsize{P} & \scriptsize{\textbf{IT2.C1.} Include only non-sensitive data in the app notification text~\cite{Mitre_ATT_CK}.}& 	\scriptsize{\textbf{DT3.R1.} The system should define limitations to the user provided data input.} & \scriptsize{P} & \scriptsize{\textbf{DT3.C1.} Setting a maximum password that can be processed~\cite{OWASP_webPage}.}\\ \hline
	
	\scriptsize{\textbf{IT2.R2.} The system shall guide users to set particular configuration settings on the mobile devices used for interaction with the system.} & \scriptsize{P} & \scriptsize{\textbf{IT2.C2.} Advise not to grant consent for device  manipulation~\cite{Mitre_ATT_CK}.} & \scriptsize{\textbf{DT3.R2.} The system should allow income traffic communications from the authorized sources.} & \multirow{2}{*}{\scriptsize{P}}& \scriptsize{\textbf{DT3.C2.} White-listening source addresses ~\cite{NistSP800_53}.}  \newline \scriptsize{\textbf{DT3.C3.} Router access control lists and firewall rules~\cite{NistSP800_53}.}\\ \hline
	\multicolumn{3}{|c||}{{\scriptsize{\textbf{IT4. Man-in-the-Middle Attack.}}}} & \multicolumn{3}{c|}{{\scriptsize{\textbf{ET1. Command Injection.}}}}\\ \hline
	
	\scriptsize{\textbf{IT4.R1.} The system should verify integrity of the transmitted data.} & \multirow{2}{*}{\scriptsize{D}} &\scriptsize{ \textbf{IT4.C1.} Cryptographic hash functions: message authentication code (MAC) algorithms~\cite{AVsecurity_AttacksDefTaxonomy_Thing_Wu}, \cite{NistSP800_53}, digital signature, and checksums~\cite{NistSP800_53}.} & \scriptsize{\textbf{ET1.R1.} The system should conduct input data validation.} & \scriptsize{C} & \scriptsize{\textbf{ET1.C1.} Server-side validation~\cite{OWASP_webPage}.} \newline\scriptsize{\textbf{ET1.C2.} Client-side validation~\cite{OWASP_webPage}}\\ \hline

	\scriptsize{\textbf{IT4.R3.} The system should ensure the confidentiality of transmitted information.}& \multirow{2}{*}{\scriptsize{P}} & \scriptsize{\textbf{IT4.C4.} Cryptographic mechanisms: SSL/TLS protocol~\cite{CISControls}, IPSec protocol suite~\cite{NistSP800_53}.}  \newline\scriptsize{\textbf{IT4.C5.}  Advanced encryption standard (AES) to encrypt wireless data in transit~\cite{CISControls}.} & \scriptsize{\textbf{ET1.R2.} The system should exploit security frameworks and libraries to validate input data before its usage.} & \multirow{2}{*}{\scriptsize{D}} & \scriptsize{\textbf{ET1.C3.} Hibernate Validator for Java~\cite{OWASP_webPage}.} \newline\scriptsize{\textbf{ET1.C4.} Data Transfer Objects for binding HTTP requests input parameters to server-side objects directly~\cite{OWASP_webPage}.} \newline\scriptsize{\textbf{ET1.C5.} Query parameterization~\cite{OWASP_webPage}.}\\ \hline

	\scriptsize{\textbf{IT4.R5.} The system should authenticate device before establishing connection.} & \scriptsize{P} & \scriptsize{\textbf{IT4.C9.} Bidirectional cryptographically based authentication~\cite{NistSP800_53}.} & \multirow{2}{2.4cm}{\scriptsize{\textbf{ET1.R4.} The organization security personnel should conduct dynamic program analysis for the launched system.}} & \multirow{2}{0.4cm}{\scriptsize{P}} & \multirow{2}{3.4cm}{\scriptsize{\textbf{ET1.C8.} Whitelisting ~\cite{FatalInjection}(e.g., SWAP, XSS-GUARD, DIDAFIT, SQLGuard).} \scriptsize{\textbf{ET1.C9.} Instruction set randomization~\cite{FatalInjection}.}
    \newline\scriptsize{\textbf{ET1.C10.} Runtime tainting~\cite{FatalInjection}.}}  \\ \cline{1-3}
	
	\scriptsize{\textbf{IT4.R6.} The system should follow the wireless capabilities policies.} & \scriptsize{P} &\scriptsize{\textbf{IT4.C10.} Usage of wireless networking capabilities only for essential functions~\cite{CISControls}, \cite{NistSP800_53}.}&&&\label{tab:SecReqsIdentification}\\\hline
\end{tabular}
\end{center}
\end{table*}

As can be seen, along with the elicited system requirements, organisational policies, and user guidelines clauses are derived. It supports the claim about the complexity of the researched scenario. Thus, the key to managing it lies in the interception of system security and human behavior management.
To illustrate how the elicited requirements can be addressed in the presented system, Table~\ref{tab:ReqsCaseAnalysis} contains examples of the concrete requirements, which mitigate risk IR5.

\begin{table}[htbp]
\caption{Security countermeasures to mitigate risk IR5}
\begin{center}
\begin{tabular}{|p{6.5cm}|p{7.5cm}|}
\hline
\bf\scriptsize{Security Requirement} &\bf \scriptsize{Security Control Components}\\
\hline
	\scriptsize{\textbf{R1.} \textit{In-Vehicle Controller} should identify unauthorized connections to the local area network.} & \scriptsize{\textbf{IT4.C3.} Authenticated application layer proxy for the network traffic that comes goes to or from the Internet~\cite{CISControls}.} \\ \hline
	
	\scriptsize{\textbf{R2.} \textit{In-Vehicle Controller} should ensure the confidentiality of transmitted via \textit{Transmission Channel} information.} & \scriptsize{\textbf{IT4.C4.} Cryptographic mechanisms: SSL/TLS protocol~\cite{CISControls}, IPSec protocol suite~\cite{NistSP800_53}.} \scriptsize{\textbf{IT4.C5.} Advanced encryption standard (AES) to encrypt wireless data in transit~\cite{CISControls}.} \\ \hline
	
    \scriptsize{\textbf{R3.} The service provider who owns the vehicles should control physical access to \textit{transmission channel} within organizational facilities.} & \scriptsize{\textbf{IT4.C6.} Wiretapping sensor~\cite{NistSP800_53}.} \scriptsize{\textbf{IT4.C7.} Locked wiring closet~\cite{NistSP800_53}.}\scriptsize{\textbf{IT4.C8.} Protection of cabling by conduit or cable trays~\cite{NistSP800_53}.}\\ \hline
	
	\scriptsize{\textbf{R4.} \textit{In-Vehicle Controller} should authenticate mobile device before establishing connection.} & \scriptsize{\textbf{IT4.C9.} Bidirectional cryptographically based authentication~\cite{NistSP800_53}.} \\ \hline
	
	\scriptsize{\textbf{R5.} \textit{Service Provider System} should follow the wireless capabilities policies.} &\scriptsize{\textbf{IT4.C10.} Usage of wireless networking capabilities only for essential functions~\cite{CISControls},\cite{NistSP800_53}.}\label{tab:ReqsCaseAnalysis}\\\hline
\end{tabular}
\end{center}
\end{table}

The provided controls should not be considered must-to-implement, but instead they give system developers and owners the understanding of possible ways to fulfil the requirements. It should be noted in the presented table, requirements are not prioritised yet, but for the usage in the system development lifecycle, the prioritisation must be done based on the earlier risk assessment, existing system architecture, and related costs.

\subsection{Preliminary Evaluation} \label{sec:Validation}

In this section, we evaluate the proposed approach of security requirements elicitation based on the developed threat model using coverage assessment criteria. The requirements comparison based on their coverage was utilised in~\cite{ahmed2014SREBP} and is based on comparison of coverage by different requirements elicitation methods. Thus, selecting an alternative method, the level of delivered security by each requirements set needs to be compared. The SQUARE~\cite{mead2007Square} and SREBP~\cite{ahmed2014SREBP} methods for eliciting security requirements can be used as alternatives. Similarly to the proposed one, these methods are applicable at the early stages of system analysis and design~\cite{pattakou2017security} but differ on the approach (risk-driven and asset-based, respectively).

After defining a set of security requirements for the Ride Fulfilment process using an alternative elicitation method, one needs to conduct a qualitative and quantitative assessment of both requirements sets. Requirements from the received sets should be categorised. Then the coverage should be assessed by analysing how many asset's attributes can be delivered by implementing the requirements. Such requirements qualitative analysis will enable the quantitative assessment of requirements categories and further comparison of the results. To avoid the carry-over effect, it is recommended to conduct requirement elicitation using an alternative method by a person unfamiliar with the results from another elicitation method. The number of elicited requirements should not be used for the assessment of the methods.

So far we have applied the SREBP method to the Ride Fulfilment process that resulted in 36 elicited security requirements. The comparison is conducted based on 11 security requirements categories~\cite{SecReqsFiresmith}: identification, authentication, authorisation, accounting, audit, non-repudiation, privacy, survivability, integrity, system maintenance, and intrusion detection. 

The coverage assessments\footnote{Set of requirements elicited usign SREBP method and the requirements coverage assessment can be accessed in \textcolor{blue}{\url{https://doi.org/10.6084/m9.figshare.14724957.v3}}.} shows that the requirements set elicited using the presented in the paper approach covers 37\% of security, while the set elicited using SREBP results in the coverage of 19\%. The elicited by SREBP requirements are asset-oriented, and therefore, more detailed in terms of system architecture. Consequently, they can be described as functional rather that system requirements. Nevertheless, the requirements sets are highly overlapping in such categories as identification, authentication, authorisation, and privacy. The SREBP method resulted in none requirements which would address system maintenance and intrusion detection, while these categories were addressed by the proposed set of requirements. One key difference is that SREBP is an asset-based approach, and therefore covers patterns of assets protection common for any business processes. In contrast, the proposed approach enables to introduce requirements which treat exactly those vulnerabilities which an attacker would target in the Passenger-AV interaction scenario. Therefore, the demonstrated in the paper elicitation method pinpoints more security aspects than SREBP.

To sum up, the set of security requirements elicited based on the developed threat model specially for the Passenger-AV interaction mitigate the more risks than the set elicited using an alternative method. The proposed set contains more general system requirements, and the comparative one - more granular but having less impact on the system overall. Therefore, we conclude that the presented on the paper requirements are valid and can be used by the developers of AV systems as a baseline for protection Passenger-AV interaction from the malicious attacks.

\section{Concluding Remarks} \label{sec:Discussion} 


To address the case analysis's main goal, we have conduct the literature review to find the methodology of information risks analysis for Passenger-AV interaction. 
We observe a knowledge gap in the autonomous vehicle technology research. Hence, to conduct case analysis, we have chosen the existing ISRM practices from the software development sphere and applied that to the case with intention to check their applicability in the context of AV system. 
Thus, we have applied the threat-driven approach to security requirements elicitation for Passenger-AV interaction. The research's main result is a developed threat model, which can be used as a baseline for threat-driven requirements elicitation. The developed model contains threats applicable for the Passenger-AV interaction in the case of ride-hailing service usage. It is assumed that an external provider delivers an infotainment system, meaning that it is not integrated into the AV. For the further usage of the presented approach, the system architecture should be considered since security risks may differ from the presented as they are assets-oriented.

The presented threat model and approach of ISRM aim to address risks related to the intentional harmful attacks, leaving out of scope unintentional threats caused by the system users. For applying the presented result to other cases, one should pay attention to the used stack of technologies in the system implementation, as there may exist several technology-specific attacks, and thus, they are not relevant to the purpose of our research.

The research is conducted using the case study analysis. This method in the autonomous vehicle field is especially prominent and applicable, as the field is relatively new and the technology is still under development. Consequently, the phenomenon is not yet determined, and the approaches for tackling it are not standardised either. Therefore, to check the applicability of the existing methods in the autonomous vehicle context, a case analysis is a fruitful method for discovering valuable insights into current theories usage and broad field. 

\paragraph{Limitations}
The further validation of the derived security requirement are needed to evaluate its coverage. We do not declare that the created threat and countermeasure models are complete, but we aim to propose them as a baseline for a systematic approach for improving autonomous vehicles' system quality for the ride-hailing service providers. The paper scope is limited by Passenger--AV interaction and focused on the application layer of the AV system. However, for ensuring information security in the on-the-move autonomous vehicle, similar research should be conducted to analyse information security risks in the other processes in AV (e.g., vehicle-to-infrastructure and vehicle-to-vehicle interactions).

\paragraph{Future Work} In our study we considered the complex system consisting of AV, an external service provider, and a human (passenger in our case). Thus the requirements for treating security risks should be defined not only for the system; but one should also assess the expectations for the companies and users to behave according to the internal security policies and user guidelines. Further research should be done to extend our result with attacks assessment, including likelihood and impact of the threats and overall assessment of the risk level. Additionally, evaluation of the derived requirements elicitation approach should be extended by comparing the requirements set with another created using an alternative elicitation method - SQUARE, coverage comparison of need to be finished.

Based on the preliminary results of requirements evaluation, we conclude that the elicited requirements are characterised by low granularity. Thus a set consists of system non-functional requirements, and organisational guidelines may be transformed into a set of more detailed requirements. Nevertheless, the proposed approach of elicitation is based on the best practices of treating attacks implementation; therefore, the elicited requirements are supported with the possible security controls which developers may use. Therefore, in future work, we will combine the demonstrated elicitation method with another to increase the level of requirements granularity.

The presented results, namely the threat and risk models also can be generalized by applying them to another case of Passenger-AV interaction, for example, in case of the other system architecture. Also, we strive to raise the discussion about the importance of implementing the security requirements for the Passenger-AV interaction for treating security and privacy issues that would increase the trust of society in general and single users to autonomous systems to ensure a high-quality user experience.


\begin{acks}
 This paper is supported in part by EU Horizon 2020 research and innovation programme under grant agreement No 830892, project SPARTA and European Social Fund via Smart Specialisation project with Bolt.
\end{acks}

\bibliographystyle{ACM-Reference-Format}
\bibliography{SRM_AV}


\end{document}